\documentclass[12pt,english]{article}
\usepackage[T1]{fontenc}
\usepackage[latin9]{inputenc}
\usepackage{geometry}
\usepackage{amsmath}
\usepackage{amsthm}
\usepackage{amssymb}
\usepackage{graphicx}

\makeatletter

\providecommand{\tabularnewline}{\\}

\numberwithin{equation}{section}
\numberwithin{figure}{section}
\newcommand{\lyxaddress}[1]{
\par {\raggedright #1
\vspace{1.4em}
\noindent\par}
}

\makeatother

\usepackage{babel}
\begin{document}

\title{Dark information of black hole radiation raised by dark energy}

\author{Yu-Han Ma$^{1,2}$, Jin-Fu Chen$^{1,2}$, Chang-Pu Sun$^{1,2}$}
\maketitle

\lyxaddress{$^{1}$Beijing Computational Science Research Center, Beijing 100193,
China\\
 $^{2}$Graduate School of Chinese Academy of Engineering Physics,
Beijing 100084, China\\
}
\begin{abstract}
The ``lost'' information of black hole through the Hawking radiation
was discovered being stored in the correlation among the non-thermally
radiated particles {[}Phys. Rev. Lett 85, 5042 (2000), Phys. Lett.
B 675, 1 (2009){]}. This correlation information, which has not yet
been proved locally observable in principle, is named by dark information.
In this paper, we systematically study the influences of dark energy
on black hole radiation, especially on the dark information. Calculating
the radiation spectrum in the existence of dark energy by the approach
of canonical typicality, which is reconfirmed by the quantum tunneling
method, we find that the dark energy will effectively lower the Hawking
temperature, and thus makes the black hole has longer life time. It
is also discovered that the non-thermal effect of the black hole radiation
is enhanced by dark energy so that the dark information of the radiation
is increased. Our observation shows that, besides the mechanical effect
(e.g., gravitational lensing effect), the dark energy rises the the
stored dark information, which could be probed by a non-local coincidence
measurement similar to the coincidence counting of the Hanbury-Brown
-Twiss experiment in quantum optics.
\end{abstract}


\section{Introduction}

Dark energy and dark matter are mysterious to compose our universe
together with various material particles \cite{69=000025,Cosmology},
which can form the generic black holes in some circumstance. While
dark matter are now widely thought to be some form of massive exotic
particles, such as primordial black hole formed within the first several
second of our universe \cite{orgin of dark matter}, it seems that
black holes have nothing to do with dark energy because dark energy
is not affected by gravity, no matter what size the black hole is
and no matter where the dark energy locates around the horizon of
the black hole. However, dark energy can causes some ``opposite gravity
effect'' so that the universe will expand faster and faster, thus
it makes everything in the universe become more distant so that the
gravitational lensing effect can be revised by dark energy \cite{dark energy metric}
. In other hand, dark energy can push the black hole horizon away
from its center in increase the horizon area (entropy) and thus, in
turn, it can exert significant influences on both the black hole radiation
and the corresponding information loss. In this paper we will investigate
these influences based on statistical mechanics about entropy . 

It is well known that, dark energy composes 69\% of our universe while
the dark matter and ordinary matter constitute the remaining 31\%
\cite{69=000025}. Many theories about dark energy have been proposed
\cite{de-1,de-2}, including a constant energy density filled in space,
or some scalar fields named quintessence \cite{de-3}. However, dark
energy is still a hypothesis used to explain the accelerating expansion
of the universe \cite{de-4}, and a convincing theory for dark energy
is lacked yet. Recently, LIGO and Virgo have made a breakthrough to
detect gravitational waves \cite{ligo-1}, the provided experimental
data showed new possible evidence for dark energy \cite{ligo-de-1,ligo-de-2,ligo-de-3,ligo-de-4}.
Besides the recently discovered gravitational lensing effect \cite{dark energy metric}
and the estimated proportion, more detailed properties of the dark
energy and its influence on other objects, such as thermodynamic influences,
have not yet been comprehensively understood and studied, both in
theory and experiment.

\begin{table}
\caption{Correction of dark energy on Black hole radiation}
\centering{}%
\begin{tabular}{lcc}
\hline 
 & Without dark energy & \multicolumn{1}{c}{With dark energy }\tabularnewline
\hline 
Hawking temperature $T_{H}$ & $\left(8\pi M\right)^{-1}$ & $\left(8\pi M\right)^{-1}\left(1-16M^{2}\Lambda/3\right)$\tabularnewline
Black hole life time $t$ & $5120\pi M^{3}$ & $5120\pi M^{3}\left(1+56M^{2}\Lambda/5\right)$\tabularnewline
Dark information $I$ & $8\pi E_{a}E_{b}$ & $8\pi E_{a}E_{b}\left(1+16M^{2}\Lambda\right)$\tabularnewline
\hline 
\end{tabular}
\end{table}

In view of the expanding of the universe, as a result of the repulsion
effect provided by dark energy, the celestial bodies in the universe
will be stretched by the expansion of the space, thus, their geometry
are changed. Let us now pay our attention to the influences of dark
energy on black hole. As a kind of celestial body, one of its geometric
features, the surface area, will be affected by the dark energy. It
follows from the black hole thermodynamics that the surface area of
a black hole, namely the horizon area, corresponds to the Bekenstein-Hawking
(B-H) entropy of black hole \cite{area law}. In addition, as we have
shown, the Hawking radiation spectrum can be straightforwardly derived
form the explicit expression of the B-H entropy with mass, charge
and angular momentum as three hair variables \cite{yhma}. Therefore,
the dark energy may influence the black hole radiation spectrum. As
a result, the change of the radiation spectrum of the black hole will
effect on various properties of the black hole and its radiation,
such as the temperature of the radiation field, namely the Hawking
temperature, the life time of the black hole and so on. 

Considering the non-thermal effect of the black hole radiation, Zhang
et al. \cite{key-QYC} found that there exists information correlation
among the radiated particles from the black hole. The two or more
particle correlation can be regarded as a kind of hidden information
of the black hole because once it were ignored, the paradox of black
hole information loss would appear. In other words, the information
we used to believe being lost in the Hawing radiation process is actually
stored as radiated particles' correlation in the radiation field \cite{ice3,ice2,ice}.
Now we recognized that the correlation information is dark since it
can not be observed locally even though the Hawking radiation can
be finally measured experimentally. Actually, to explore quantum or
classical correlation, two or more non-local probes are needed to
make a coincidence measurement similar to the coincidence counting
of the Hanbury-Brown -Twiss experiment in quantum optics \cite{QO}.
In this paper we are interested in whether dark energy will increase
or reduce the dark information characterized by the mutual entropy. 

All in all, the study of the black hole radiation with dark energy
will enlighten us to comprehend the influences of dark energy on the
thermodynamic properties of black hole and its ``lost'' information.
To this end , we first derive the correction to the radiation spectrum
of black hole with the existence of dark energy. The universal approach
based on the canonical typicality \cite{key-CT1,key-CT2,key-CT3}
to achieve the black hole radiation spectrum is confirmed by the quantum
tunneling approach with the Schwarzschild black hole as an example.
According to the corrected radiation spectrum with dark energy, the
Hawking temperature, the life time of black hole, and the dark information
in the black hole radiation are obtained for the Schwarzschild black
hole of mass $M$ with non-vanishing cosmological constant $\Lambda$
and the state parameter $w$ \cite{dark energy metric}. 

The main results of this paper are illustrated in Tab. 1, where $E_{a}$and
$E_{b}$ are the energy of two radiated particles $a$ and $b$, respectively;
the natural units with $\hbar=G=k_{B}=c=1$ are adopted thereafter
for simplicity. It should be mentioned here that the results shown
in Tab. 1 are hold only in the cosmological constant model, where
the state parameter $w=-1$. For an arbitrary $w$, the corresponding
results are more complicated and will be demonstrated in the main
text of this paper as Eqs. (\ref{eq:T}), (\ref{eq:t}), and (\ref{eq:I}).
Our results show that the dark energy makes the Hawking radiation
colder, the life time of the black hole longer, and raises more dark
information of Hawking radiation.

This paper is organized as follows. In Sec. 2, we first review the
canonical typicality based general approach we developed to derive
the radiation spectrum of the black hole. Then, we briefly introduce
the quantum tunneling method for obtaining the radiation spectrum
dynamically. In Sec. 3, we calculate the horizon radius of a Schwarzschild
black hole by taking account of the effect of dark energy. Then, we
use it to obtain the radiation spectrum from our statistical mechanical
approach. We also make a verification of this spectrum with the quantum
tunneling method. Furthermore, according to the corrected radiation
spectrum, we consider the black hole's evaporation and give the corrected
Hawking temperature and black hole life time in Sec. 4. In Sec. 5,
we study the influence of dark energy on dark information among Hawing
radiation. We will conclude and make some discussions in Sec. 6. In
Appendix, we provide the detailed derivation of the canonical typicality
-based approach. 

\section{General approach to radiation spectrum of black hole }

The radiation spectrum of black hole describes the statistical distribution
of particle's energy after the particles cross the horizon through
the Hawking process. From the microscopic point of view, Hawking process
is composed of the following two steps: (i) quantum fluctuations make
a large number of particle pairs created and annihilated near event
horizon; (ii) while a positive energy particle from a pair created
virtually inside the horizon escapes out of the horizon through tunneling,
the Hawking radiation happens. This means that the Hawking process
is a dynamic process. Thus, it seems that the dynamic analysis of
radiated particles is needed to derive the Hawking radiation spectrum.
With the help of the curved space-time quantum field theory, the radiation
spectrum of black hole was first obtained by Hawking, and it was found
to obey the thermal distribution \cite{key-sw1,key-sw2}. 

In fact, this thermal spectrum implies that the entropy will increase
through the Hawking process, which leads to the black hole information
paradox. In order to resolve this paradox, different schemes have
been put forward \cite{il1,il2,il3,il4,il5,il6,il7,il8,il9,il191},
such as modifying the radiation spectrum itself \cite{il6}. When
the constraint of energy conservation is considered, the black hole
radiation spectrum is shown to be not perfectly thermal. This non-thermal
radiation spectrum have been proved to satisfy the requirement of
information conservation \cite{key-QYC}. 

\subsection{Canonical typicality based approach}

In our recent study \cite{yhma}, we obtained the non-thermal black
hole radiation spectrum by a purely statistical mechanical approach
based on canonical typicality. It is emphasized that this approach
do not refer to any dynamics of particle tunneling. In this section,
we will briefly review this general approach (for the details please
see Appendix). It follows from the canonical typicality \cite{yhma,key-CT1,key-CT2,key-CT3}
or the micro-canonical hypothesis that the density matrix of an arbitrary
black hole B with three ``hairs'', namely mass $M$, charge $Q$,
and angular momentum $J$, reads

\begin{equation}
\rho_{\textrm{B}}=\sum_{i}\frac{1}{\Omega_{\textrm{B}}\left(M,Q,J\right)}\left|M,Q,J\right\rangle _{i}\left\langle M,Q,J\right|.
\end{equation}
Here, $\left|M,Q,J\right\rangle _{i}$ is the $i$th eigenstate of
B, and $\Omega_{B}\left(M,Q,J\right)$ is B's number of microstate.
When the black hole evaporates, the system evolves into two components,
the radiation field R and the remaining black hole $\textrm{B}'$.
By taking into consideration the fact that the black hole ``hairs''
are conserved quantities, then tracing over all the degree of freedom
of $\textrm{B}'$, the density matrix of R is obtained as

\begin{equation}
\rho_{\textrm{R}}=\sum_{\omega,q,j}p\left(\omega,q,j,M,Q,J\right)\left|\omega,q,j\right\rangle \left\langle \omega,q,j\right|,\label{eq:rouRgeneral}
\end{equation}
where $\left|\omega,q,j\right\rangle $ is the eigenstate of R, which
has mass $\omega$, charge $q$, and angular momentum $j$. The distribution
probability of the radiation is given by

\begin{equation}
p\left(\omega,q,j,M,Q,J\right)=\textrm{e}^{-\Delta S_{\textrm{BB}'}\left(\omega,q,j,M,Q,J\right)},\label{eq:densityRa}
\end{equation}
with

\begin{equation}
\Delta S_{\textrm{BB}'}\left(\omega,q,j,M,Q,J\right)=S_{\textrm{B}}\left(M,Q,J\right)-S_{\textrm{B}'}\left(M-\omega,Q-q,J-j\right)\label{eq:entropy difference}
\end{equation}
being the entropy difference between B and $\textrm{B}'$. Equations
(\ref{eq:rouRgeneral}-\ref{eq:entropy difference}) offer us a straightforward
approach to calculate the radiation spectrum of black hole. Only when
the entropy linearly depends on energy, the spectrum will be perfectly
thermal. In fact, in the thermodynamic limit, the higher orders of
energy in the distribution probability are always ignored for a large
black hole, thus the thermal equilibrium distribution appears. However,
when the system being studied is not large enough, the higher-order
terms of energy in the distribution probability need to be preserved.
Then the emergent non-thermal spectrum means the correlation inside
the system due to the energy (charge, angular momentum) conservation
law. It follows from this observation that we have found the non-thermal
distributions for some specific systems \cite{key-29,key-28}.

It can be seen from Eq. (\ref{eq:densityRa}) that the radiation spectrum
can be given from the entropy of black hole, the Bekenstein-Hawking
(B-H) entropy. According to so-called black hole area law \cite{area law},
the B-H entropy is proportional to the area of horizon. Without loss
of generality, we can express the black hole entropy as, , 

\begin{equation}
S_{\textrm{B}}\left(M,Q,J\right)=S_{BH}\left(M,Q,J\right)=\frac{A_{H}\left(M,Q,J\right)}{4}=\pi R_{H}^{2}\left(M,Q,J\right),\label{eq:entropy}
\end{equation}
where, $A_{H}\left(M,Q,J\right)$ and $R_{H}\left(M,Q,J\right)$ are
the area and radius of the horizon, respectively. With Eqs. (\ref{eq:densityRa})
and (\ref{eq:entropy}), we re-write the radiation spectrum as the
function of horizon radius as

\begin{equation}
p\left(\omega,q,j,M,Q,J\right)=\textrm{e}^{-\pi\left[R_{H}^{2}\left(M,Q,J\right)-R_{H}^{2}\left(M-\omega,Q-q,J-j\right)\right]}.\label{eq:pr}
\end{equation}
As an illustration, we apply Eq. (\ref{eq:pr}) to a Schwarzschild
black hole with horizon radius being $R_{H}=2M$. We then straightforward
obtain the distribution probability for a particle with energy $\omega$
as

\begin{equation}
p\left(\omega,M\right)=\exp\left[4\pi\left(M-\omega\right)^{2}-4\pi M^{2}\right]=\exp\left[-8\pi\omega\left(M-\frac{\omega}{2}\right)\right].\label{eq:pr-sbh}
\end{equation}
This non-thermal spectrum is exactly the same as Eq. (10) of Ref \cite{il6},

\begin{equation}
\Gamma=\textrm{e}^{-8\pi\omega\left(M-\omega/2\right)},
\end{equation}
which even was derived through the quantum tunneling perspective. 

Obviously, when the mass of a black hole is large, i.e. , $M\rightarrow\infty$,
Eq. (\ref{eq:pr-sbh}) can be approximated as

\begin{equation}
p\left(\omega,M\right)\approx\exp\left(-8\pi M\omega\right),
\end{equation}
This is the original thermal radiation spectrum obtained by Hawking.
Therefore, we can say that we have developed an effective method to
calculate the non-thermal radiation spectrum of the black holes, and
it does not need to analysis the tunneling dynamics for the particle
crossing the horizon. Generally, to find the black hole radiation
spectrum, one can only derive the horizon radius by analyzing the
singularity of the metric for a given black hole at first. Secondly,
by making use of Eq. (\ref{eq:pr}), the specific form of the radiation
spectrum is obtained. 

\subsection{Quantum tunneling based approach}

In the aforementioned derivation, we only make use of the statistical
properties of the black hole. This method obtains the black hole radiation
spectrum without referring the particle's dynamics , thus is only
of the statistical mechanics. Another approach obtaining black hole
radiation spectrum is the quantum tunneling method, which was first
introduced by Parikh and Wilczek to derive the non-thermal radiation
spectrum of Schwarzschild black hole \cite{il6}. This method was
even used for different types of black holes to obtain their radiation
spectrum successfully \cite{key-18,key-34,key-35,key-36,key-37,key-38,key-39,key-40,key-41,key-42,key-43}.
These spectra have been proved to meet the requirement of information
conservation, and are exactly consistent with our result of Eq. (\ref{eq:entropy})
obtained statistical-mechanically. 

We now brief introduce this tunneling approach. Since there is a strong
gravitational potential near the horizon, the particles escape from
the horizon through the Hawking process can be considered as a quantum
tunneling process. With the semi-classical approximation, the tunneling
probability for the particle can be written as \cite{il6}

\begin{equation}
\Gamma=\exp\left[-2\textrm{Im}(I)\right],
\end{equation}
where

\begin{equation}
\textrm{Im}(I)=\int_{r_{in}}^{r_{out}}p_{r}dr
\end{equation}
is the imaginary part of the action $I$ for a positive energy particle
tunneling outside crossing the horizon. $r_{in}$ ($r_{out}$) is
the horizon radius before (after) the particle escaping from the horizon,
and $p_{r}$ is the radial momentum of the particle. Making use of
the Hamilton\textquoteright s equation

\begin{equation}
\dot{r}=\frac{dH}{dp_{r}},
\end{equation}
we can eliminate the momentum by energy in the action as 

\begin{equation}
\mathrm{Im}\left(I\right)=\mathrm{Im}\int_{H_{i}}^{H_{f}}\int_{r_{in}}^{r_{out}}\frac{dr}{\dot{r}}dH.\label{eq:ImI}
\end{equation}
Here, $\dot{r}$ is the radial null geodesics, and it can be obtained
from the black hole metric, and $H_{i}$ ($H_{f}$) is the energy
of black hole before (after) the particle escaping from the horizon.
With the specific metric form of a given black hole, we can derive
the corresponding radial null geodesics $\dot{r}$ , then we finish
the integral in Eq. (\ref{eq:ImI}) and thus the tunneling probability
is obtained. The detailed calculation of this approach in the case
with dark energy is presented in the Sec. 3.2. 

Moreover, the statistical mechanical approach allows us to calculate
the distribution probability in the radiation process without analyzing
the dynamics of particle tunneling. The obtained results are exactly
the same as that given by the tunneling approach. This indicates that
the statistical mechanical properties of the radiation spectrum may
have an intrinsic relationship with the tunneling dynamics of the
particles through the Hawing process. Some relevant studies also suggest
this, for example, the number of micro-state for black holes can be
calculated directly from the tunneling spectrum \cite{key-31}, and
the radiation spectra of the black holes have been proved to be independent
of the black holes' geometry \cite{key-30}.

\section{Black hole radiation spectrum with dark energy }

In this section, we are going to derive the radiation spectrum of
the Schwarzschild black hole in the existence of the dark energy from
both the statistical mechanical approach and the tunneling approach.
To obtain the specific form of the spectrum, both of these two methods
need to express the horizon radius as the function of the black hole
mass. Therefore, we first derive the horizon radius from the metric
of the black hole. When the effect of the dark energy is taken into
account, the line element for the Schwarzschild black hole reads \cite{dark energy metric,metric-r}

\begin{equation}
\textrm{d}s^{2}=\left[1-\frac{2M}{r}-2\left(\frac{r_{o}}{r}\right)^{3w+1}\right]\textrm{d}t^{2}-\frac{\textrm{d}r^{2}}{\left[1-\frac{2M}{r}-2\left(\frac{r_{o}}{r}\right)^{3w+1}\right]}-r^{2}\left(\textrm{d}\theta^{2}+\sin^{2}\theta\textrm{d}\phi^{2}\right),\label{eq:lineelements}
\end{equation}
where $M$ is the mass of the black hole. $r_{o}$ is a scale factor,
and it can be represented by the cosmological constant $\Lambda$
as $r_{o}=\sqrt{6/\Lambda}$. The state parameter $w$ is a constant
within the range $-1<w<-1/3$. According to the definition of the
black hole horizon, its radius satisfies the following equation

\begin{equation}
1-\frac{2M}{R_{H}}-2\left(\frac{r_{o}}{R_{H}}\right)^{3w+1}=1-\frac{2M}{r}-2\left(\frac{6}{\Lambda R_{H}^{2}}\right)^{\frac{3w+1}{2}}=0.\label{eq:rh}
\end{equation}
Assuming that the dark energy affects the geometry of the black hole
slightly, we can solve Eq. (\ref{eq:rh}) perturbatively by taking
$R_{H}=R_{H}^{0}+\delta$. Here, $\delta/R_{H}^{0}\ll1$, and $R_{H}^{0}=2M$
is the horizon radius of the Schwarzschild black hole in the case
that the dark energy does not exist. Then we have

\begin{equation}
\frac{\delta}{2M}=2\left(2M\right)^{2\xi}\left(\frac{\Lambda}{6}\right)^{\xi},
\end{equation}
where $\xi\equiv-(3w+1)/2\in(0,1)$ is a re-defined positive constant.
Finally, the horizon radius of the black hole is obtained as

\begin{equation}
R_{H}=2M+4M^{2\xi+1}\left(\frac{2\Lambda}{3}\right)^{\xi}\equiv2M+4M^{2\xi+1}f\label{eq:R}
\end{equation}
with

\begin{equation}
f=f\left(\xi,\Lambda\right)=\left(\frac{2\Lambda}{3}\right)^{\xi},
\end{equation}
where we have only kept the first order of $f$. It is seen from Eq.
(\ref{eq:R}) that the dark energy will increase the horizon radius.
In other words, the black hole becomes bigger in the existence of
dark energy. This can be easily understood from the repulsion effect
of dark energy to the universe. In comparison with the attraction
effect of the gravitation among normal matters, dark energy produces
the repulsive effect, which is considered as the reason for the expansion
of the universe. Our result shows that the black hole is indeed extended
by the dark energy. The quantitative dependence description of this
extension effect is given in Eq (\ref{eq:R}). 

\subsection{Radiation spectrum from statistical mechanical approach }

From the correction of the dark energy to the horizon radius, we will
further study the influences of the dark energy on the black hole
radiation. Through the statistical mechanical method we introduced
in Sec. 2.1, we substitute the horizon radius {[}Eq. (\ref{eq:R}){]}
into Eq. (\ref{eq:pr}), and then the corrected radiation spectrum
of Schwarzschild black hole is obtained as

\begin{equation}
p\left(\omega,M\right)=\textrm{exp}\left\{ -8\pi M\left[1+4\left(\xi+1\right)M^{2\xi}f\right]\omega+4\pi\left[1+4\left(\xi+1\right)\left(2\xi+1\right)M^{2\xi}f\right]\omega^{2}\right\} ,\label{eq:p}
\end{equation}
where we have kept the second order of $\omega$, and the higher orders
$O\left(\omega^{2}\right)$ are ignored. This indicates that the radiation
spectrum or the tunneling probability is $\xi$ and $\Lambda$ -dependent.
In the case of the cosmological constant model of the dark energy,
we have $\xi=1$ with the state parameter $w=-1$, and Eq. (\ref{eq:p})
reduces to a more concise form as

\begin{equation}
p\left(\omega,M\right)=\exp\left[-8\pi M(1+8M^{2}f)\omega+4\pi\left(1+24M^{2}f\right)\omega^{2}\right].\label{eq:p1}
\end{equation}
If the effect of the dark energy approaches 0, the above probability
would reduce to $p\left(\omega,M\right)=\exp\left[-8\pi\omega\left(M-\omega/2\right)\right]$,
which is exactly consistent with the Parikh -Wilczek (P-W) spectrum.
We then plot Eq. (\ref{eq:p1}) in Fig. 1, where the black hole mass
$M=1$, and the P-W spectrum is also potted as a comparison. 
\begin{figure}
\begin{centering}
\includegraphics[width=10cm]{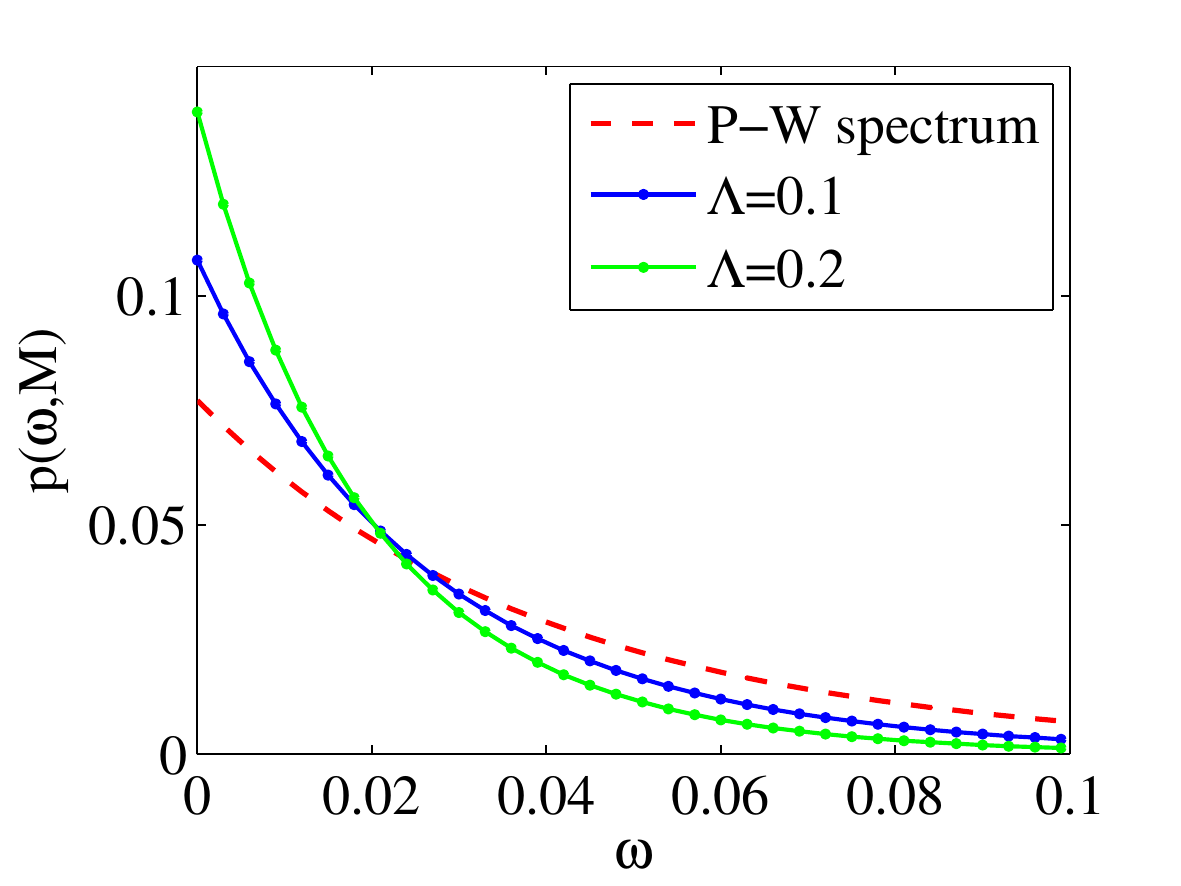}
\par\end{centering}
Figure. 1 (Color online). Radiation distribution probability as the
function of the particle energy. The red dash line is the Parikh -Wilczek
spectrum, while the blue dotted line (green dotted line) represents
the radiation spectrum with the dark energy, where the corresponding
cosmological constant is taken as $\Lambda=0.1$ ($\Lambda=0.2$).
\end{figure}
 It can be seen from Fig. 1 that the radiation spectrum is sharply
affected by the dark energy. As the influence of the dark energy becomes
lager ($\Lambda$ increasing ), the distribution of the low energy
parts in the radiation increases, and the proportion of high energy
radiation decreases. This implies that the temperature of the radiation
becomes lower due to the existence of the dark energy, which agrees
with the direct theoretical analysis in Sec. 4.

\subsection{The verification from quantum tunneling method }

To confirm the result obtained through our statistical mechanical
approach, we calculate the radiation spectrum by utilizing the tunneling
approach in the case with the dark energy. In the quantum tunneling
method, we need to analyze the dynamic behavior of the particle when
it is crossing the horizon. Unfortunately, the Schwarzschild coordinates,
as we introduced in Eq. (\ref{eq:lineelements}), are singular at
the horizon. Thus, it is necessary to choose a new coordinate system
to eliminate this singularity at the horizon. One of the suitable
choice, first given by Painlevé \cite{key-44,key-45}, is to introduce
a time coordinate, in our case, as follows

\begin{equation}
dt=dt_{s}+\frac{\sqrt{2M+r_{o}\left(\frac{r_{o}}{r}\right){}^{3w}}\sqrt{r}}{2M+r_{o}\left(\frac{r_{0}}{r}\right){}^{3w}-r}dr,
\end{equation}
where $t_{s}$ is the Schwarzschild time. According to Eq. (\ref{eq:lineelements}),
the line element is re-written in the new coordinates as

\begin{equation}
ds^{2}=-\left[1-\frac{2M}{r}-\left(\frac{r_{o}}{r}\right)^{3w+1}\right]dt^{2}+2\sqrt{\frac{2M}{r}+\left(\frac{r_{o}}{r}\right){}^{3w+1}}drdt+dr^{2}+r^{2}d\Omega^{2}.\label{eq:new}
\end{equation}
The above equation shows that there is no singularity at the horizon.
Then, from Eq. (\ref{eq:new}), we obtain the radial null geodesics

\begin{equation}
\dot{r}\equiv\frac{dr}{dt}=\pm1-\sqrt{\frac{2M}{r}+\left(\frac{r_{o}}{r}\right)^{3w+1}},\label{eq:rdot}
\end{equation}
where the upper (lower) sign is corresponding to the outgoing (ingoing)
geodesics. Consequently, the imaginary part of the action $I$ for
an s-wave outgoing positive energy particle crossing the horizon can
be expressed as

\begin{equation}
\mathrm{Im}\left(I\right)=\mathrm{Im}\int_{M}^{M-\omega}\int_{r_{in}}^{r_{out}}\frac{dr}{1-\sqrt{\frac{2M^{\prime}}{r}+\left(\frac{r_{o}}{r}\right)^{3w+1}}}\left(dM^{\prime}\right),\label{eq:intergral}
\end{equation}
where we have used the replacement $dH\rightarrow dM^{\prime}$. $r_{in}$
($r_{out}$ ) is the horizon radius before (after) the Hawking process
corresponding to the particle with energy $\omega$, and can be derived
from Eq. (\ref{eq:R}). Letting
\begin{equation}
x(M^{\prime})=\sqrt{\frac{2M^{\prime}}{r}+\left(\frac{r_{o}}{r}\right)^{3w+1}}
\end{equation}
and exchanging the order of the integral in Eq. (\ref{eq:intergral}),
we have
\begin{equation}
\mathrm{Im}\left(I\right)=\mathrm{Im}\int_{r_{in}}^{r_{out}}\int_{x_{i}}^{x_{f}}\frac{xr}{1-x}dxdr,
\end{equation}
where $x^{i}=x\left(M\right)$, and $x^{f}=x\left(M-\omega\right)$.
Note that $x_{i}<1<x_{f}$, so that the integral contain a singular
point $x=1$. With the contour evaluated via the prescription $x\rightarrow x-i\epsilon$,
we have the imaginary part of the action
\begin{align}
\mathrm{Im}\left(I\right) & =\int_{r_{in}}^{r_{out}}-\pi rdr=\frac{1}{2}\pi\left(r_{in}^{2}-r_{out}^{2}\right)\\
 & =2\pi\omega^{2}-4\pi M\omega-16\pi\left(\xi+1\right)M^{2\xi+1}f\omega+8\pi\left(\xi+1\right)\left(2\xi+1\right)M^{2\xi}f\omega^{2}.
\end{align}
 As an ingoing negative energy particle contributes the same value,
we obtain 

\begin{equation}
\Gamma=\textrm{e}^{-2\mathrm{Im}\left(I\right)}=\textrm{exp}\left\{ -8\pi M\left[1+4\left(\xi+1\right)M^{2\xi}f\right]\omega+4\pi\left[1+4\left(\xi+1\right)\left(2\xi+1\right)M^{2\xi}f\right]\omega^{2}\right\} ,\label{eq:Gramma}
\end{equation}
which exactly agrees with Eq. (\ref{eq:densityRa}) from our statistical
mechanical approach.

Until now, the Black hole radiation spectrum have been obtained from
the perspective of the statistical physics and the quantum tunneling,
respectively. As shown in Eqs.(\ref{eq:p}) and (\ref{eq:Gramma}),
these two methods provide the same result, indicating that they are
mutually corroborated. Besides, we can clearly perceive that using
the statistical mechanical way to get the black hole radiation spectrum
is much more succinct than using the tunneling method, while the latter
one need to analyze the particle's dynamics in the Hawking process.

\section{Evaporation of the Schwarzschild black hole with dark energy }

In this section, with the dark energy based radiation spectrum, we
will study the evaporation process of the black hole. Specifically,
as the two characteristics of the black hole evaporation, the Hawking
temperature and the black hole life time will be re-investigated with
the correction of the dark energy. 

The Hawking radiation \cite{key-sw1,key-sw2} implies that the black
hole has temperature. Using the curved space-time quantum field theory,
Hawking obtained the thermal radiation spectrum of the black hole.
Therefore, he explained that the temperature of the black hole i.e.,
$T_{H}=1/(8\pi M)$, is the true temperature. This is the so-called
Hawking temperature, and was first introduced according to the similarity
between the black hole laws and the thermodynamics laws. For a black
hole with a large mass, i.e., $M\gg1$, the non-thermal part of the
radiation spectrum can be ignored. Thus, from Eqs. (\ref{eq:densityRa})
and (\ref{eq:p}), we have the approximated thermal distribution probability

\begin{equation}
p\left(\omega,M\right)=\exp\left\{ -8\pi M\left[1+4\left(\xi+1\right)M^{2\xi}f\right]\omega\right\} .\label{eq:thermal}
\end{equation}
When the effect of the dark energy vanishes, namely $f\rightarrow0$,
the above distribution will reduce to the Hawking radiation spectrum 

\begin{equation}
p\left(\omega,M\right)=\textrm{e}^{-8\pi M\omega},
\end{equation}

\subsection{Hawking temperature}

It is directly obtained from Eq. (\ref{eq:thermal}) that the inverse
radiation temperature

\begin{equation}
\beta_{H}=8\pi M\left[1+4\left(\xi+1\right)M^{2\xi}f\right],\label{eq:beta}
\end{equation}
thus the Hawking temperature with the existence of dark energy reads

\begin{equation}
T_{H}=\frac{1}{8\pi M}\left[1-4\left(\xi+1\right)M^{2\xi}\left(\frac{2\Lambda}{3}\right)^{\xi}\right],\label{eq:T}
\end{equation}
where we only kept the first order of $f$ . By introducing a the
modification factor

\begin{equation}
\lambda=4\left(\xi+1\right)M^{2\xi}\left(\frac{2\Lambda}{3}\right)^{\xi},
\end{equation}
the Hawking temperature can be further simplified as

\begin{equation}
T_{H}=T_{H}^{0}(1-\lambda),\label{eq:HT}
\end{equation}
where $T_{H}^{0}=1/\left(8\pi M\right)$ is the Hawking temperature
without the dark energy. Equation (\ref{eq:beta}) shows that the
dark energy makes the Hawking radiation colder. 

In Fig. 2, we show that the result of Eq. (\ref{eq:HT})

\begin{figure}
\begin{centering}
\includegraphics[width=12cm]{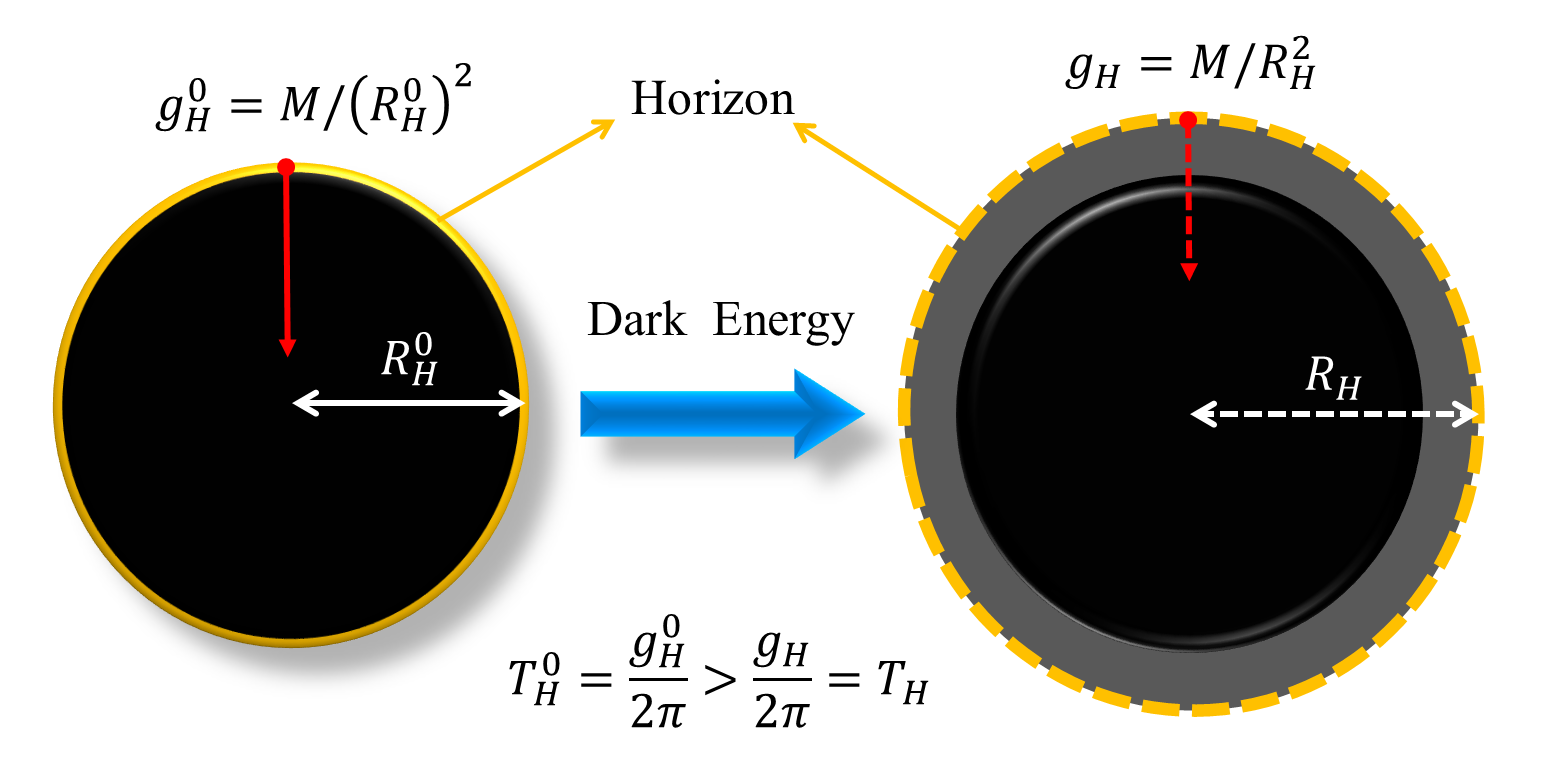}
\par\end{centering}
Figure. 2 (Color online). The dark energy based cooling mechanism
(DECM) for the black hole. The black circular areas represent the
black hole of mass $M$, $R_{H}$ and $g_{H}$ ($R_{H}^{0}$ and $g_{H}^{0}$
) are the radius and surface gravity of the black hole with (without)
dark energy, respectively. $T_{H}$ ($T_{H}^{0}$ ) is the Hawking
temperature with (without) dark energy. The repulsion effect provided
by dark energy increases the radius of the black hole, so that the
surface gravity of the black hole decreases. In the black hole thermodynamics,
the Hawking temperature is linearly dependent on the black hole surface
gravity, and is thus decreased by dark energy. 

\end{figure}
can also be understood from the black hole thermodynamics, where the
temperature of the black hole is positively correlated to the gravity
on black hole's surface. As we have shown in Sec. 3 that black hole
can be extended by the dark energy, i.e. , $R_{H}\uparrow$, leading
to the black hole surface gravity $g_{H}=M/R_{H}^{2}$ decreases,
i.e., $g_{H}\downarrow$. Therefore, the Hawing temperature $T_{H}=g_{H}/\left(2\pi\right)$
becomes lower, namely $T_{H}\downarrow$. The phenomenon that the
dark energy decreases the Hawking temperature through expanding the
space is named by the dark energy based cooling mechanism (DECM). 

In addition, the modification factor $\lambda$ decreases with $M$,
meaning that the influence of dark energy on the Hawking temperature
increases with the black hole mass. In the cosmological constant model
of the dark energy, we have the state parameter $w=-1$ and $\xi=1$.
The radiation temperature of Eq. (\ref{eq:T}) becomes 

\begin{equation}
T_{H}=\frac{1}{8\pi M}\left(1-\frac{16M^{2}\Lambda}{3}\right).
\end{equation}

Until now, we have obtained the Hawking radiation temperature with
the existence of dark energy. According to the theory of black-body
radiation, the lower the temperature is, the smaller the radiation
power is. Intuitively, when a black hole is completely radiated, the
time for this process will be longer while the total energy is a constant.
Next, we will focus on the influence of the dark energy on the life
time of the black hole. 

\subsection{Life time of the black hole}

It follows from the Stefan\textendash Boltzmann power law that the
radiation power of the black hole with temperature $T_{H}$ is

\begin{equation}
P=A_{H}\sigma T_{H}^{4}.\label{eq:P}
\end{equation}
Here, $A_{H}=4\pi R_{H}^{2}$ is the area of the horizon, and $\sigma=\pi^{2}/60$
is the Stefan constant. Using the energy conservation law for the
black hole and its radiation, one has

\begin{equation}
\frac{dM}{dt}+P=0.
\end{equation}
With the help of Eqs. (\ref{eq:beta}) and (\ref{eq:P}), the above
equation is explicitly expressed as

\begin{equation}
-\frac{dM}{dt}=4\pi\left(2M+4M^{2\xi+1}f\right)^{2}\left(\frac{\pi^{2}}{60}\right)\left\{ \frac{1}{8\pi M}\left[1-4\left(\xi+1\right)M^{2\xi}f\right]\right\} ^{4}.\label{eq:dm/dt}
\end{equation}
Eq. (\ref{eq:dm/dt}) is simplified as

\begin{equation}
-\frac{dM}{dt}=\frac{1}{15360\pi M^{2}}\left[1-\left(16\xi+12\right)M^{2\xi}f\right],
\end{equation}
by keeping the first order of $f$. Thus, the time for the black hole
evaporating from mass $M$ to $M\left(t\right)$ is

\begin{equation}
t=-\int_{M}^{M\left(t\right)}15360\pi M^{2}\left[1+\left(16\xi+12\right)M^{2\xi}f\right]dM.
\end{equation}
When $M\left(t\right)=0$, the black hole evaporates all its energy,
and the evaporation time in this case is just the life time of the
black hole, namely

\begin{equation}
t=t_{0}\left[1+\frac{3\left(16\xi+12\right)}{2\xi+3}\left(\frac{2\Lambda}{3}\right)^{\xi}M^{2\xi}\right].\label{eq:t}
\end{equation}
Here, $t_{0}=5120\pi M^{3}$ is the life time for the Schwarzschild
black hole in the absence of the dark energy. Equation (\ref{eq:t})
tells us that the dark energy makes the black hole's life time longer.
This can be seen as a reflection of the cooling effect that the dark
energy on the Hawking radiation. When the radiation temperature drops,
i.e. , $T_{H}\downarrow$, the power of the black hole radiation process
decreases, i.e. , $P\downarrow$, meaning that the energy released
per unit time becomes less. Therefore, it takes more time for the
black hole to release all its energy, namely $t\uparrow$. In the
cosmological constant model of the dark energy, we have Eq. (\ref{eq:t})
been reduced to

\begin{equation}
t=5120\pi M^{3}\left(1+\frac{56}{5}M^{2}\Lambda\right).
\end{equation}
It is found that the dark energy behaves like a ``refrigerator''
placed around the black hole, which will make the radiation of the
black hole colder and slow down the Hawing radiation process. Consequently,
the black hole's life time is prolonged by the dark energy.

\section{Dark information added by dark energy }

As the black hole evaporates so that it becomes smaller and smaller,
the non-thermal effect of its radiation increases gradually. The non-thermal
radiation has been proved to be the origin of the information correlation
\cite{key-QYC} between the emissions that being radiated out from
the black hole's horizon. In this section, we will study the influences
of the dark energy on this correlation. 

\begin{figure}
\begin{centering}
\includegraphics[width=12cm]{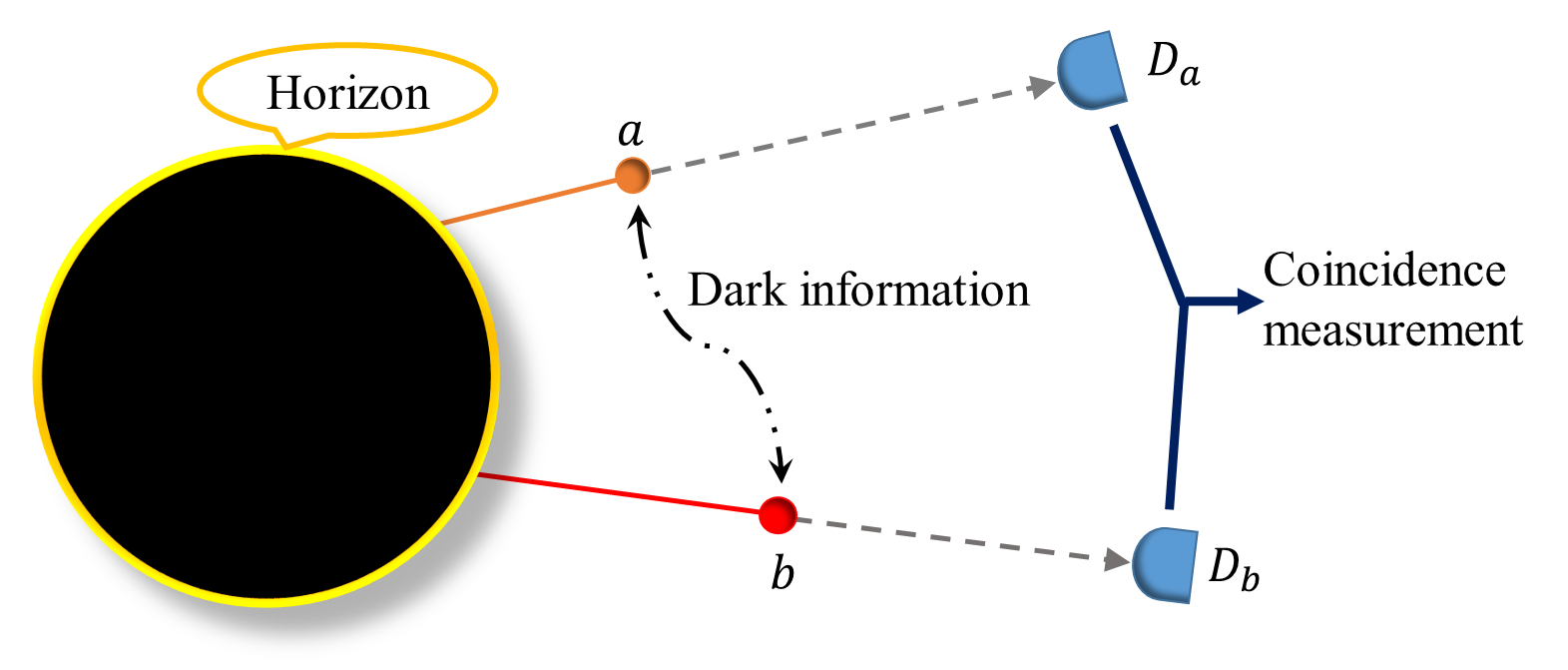}
\par\end{centering}
Figure. 3 (Color online). Probing the dark information of the black
hole radiation. Two radiated particles $a$ and $b$ escape from the
black hole horizon. The energy distribution of these two particles
are not independent of each other due to the non-thermal radiation
spectrum, implying that there exist correlation between them. Because
the particles may be radiated out from any position of the horizon,
and the directions of their momentum are random, it is almost impossible
to detect them at the same location. Suppose we have two detectors
$D_{a}$ and $D_{b}$ to detect these two particles separately in
the space. If one only focus on $D_{a}$ ($D_{b}$ ), the result of
its detection is an average over the distribution of $b$ ($a$),
thus the correlation between $a$ and $b$ is hidden and can not be
observed. To show this correlation information, the coincidence measurement
should be taken among $a$ and $b$ through the two detectors. Therefore,
we can conclude that this correlation can not be probed through local
measurement. Namely, the information is dark.
\end{figure}
As shown in Fig. 3., this information correlation can not be observed
locally even though the Hawking radiation can be measured experimentally.
In this sense the information stored in the correlation is named by
dark information.When the dark information is taken into account,
the sum of the information of the radiation field and the remaining
black hole gives the total information of the black hole system, and
thus is conserved \cite{yhma,key-QYC,key-30} in the Hawking radiation
process. Therefore, the dark information caused by the non-canonical
statistic behavior of the Hawking radiation will results in a possible
solution to the black hole information paradox. Besides, the dark
information of Hawking depends on the radiation spectrum of the black
hole, and more accurately, it depends on the specific form of the
non-thermal part of the radiation spectrum. From the distribution
probability given in Eq. (\ref{eq:p}) we can clearly see that the
existence of the dark energy changes the radiation spectrum of the
black hole, and it may also affect the dark information of the radiation.
In this section, we consider the influence of the dark energy on the
dark information, and show that the dark energy will increase the
dark information of black hole radiation. 

For the two non-independent events $a$ and $b$, they are correlated
with each other, or in other words, there exists a information correlation
between them. From the point of view of the information theory, this
correlation can be described quantitatively with the mutual information
\cite{key-mi}

\begin{equation}
I\left(a,b\right)=\sum_{a,b}p_{a,b}\ln\left(\frac{p_{a,b}}{p_{a}p_{b}}\right),\label{eq:I-1}
\end{equation}
where $p_{i}$ is the probability for event $i$ ($i=a,b$), and $p_{a,b}$
is the joint probability of these two events. Obviously, when $a$
and $b$ are independent with each other, we have $p_{a,b}=p_{a}p_{b}$.
Thus, the mutual information vanishes , $I\left(a,b\right)=0$. Because
the original radiation spectrum of the black hole discovered by Hawking
is perfectly thermal, one can easily check that the correlation among
radiations is zero. This implies that there is no dark information
in the thermal black hole radiation. 

To find the correlation among the non-thermal black hole radiations,
we choose the two events as the radiation process of the two particles
$a,b$ with energy $\omega_{a}$ and $\omega_{b}$, respectively.
In this case, the probability for each particle is given by Eq. (\ref{eq:p})
as $p_{i}=p\left(\omega_{i},M\right)$($i=a,b$), and the joint probability
$p_{a,b}=p\left(\omega_{a}+\omega_{b},M\right)$. Substituting Eq.
(\ref{eq:p}) into Eq. (\ref{eq:I-1}), we obtain the dark information
as (for detailed calculation, please see Appendix B)

\begin{equation}
I\left(a,b\right)=8\pi\left[1+4\left(\xi+1\right)\left(2\xi+1\right)M^{2\xi}f\right]E_{a}E_{b},\label{eq:I}
\end{equation}
which can be further simplified as

\begin{equation}
I\left(a,b\right)=8\pi\left(1+\mu\right)E_{a}E_{b},
\end{equation}
where 

\begin{equation}
\mu=4\left(\xi+1\right)\left(2\xi+1\right)M^{2\xi}\left(\frac{2\Lambda}{3}\right)^{\xi}
\end{equation}
is a positive modification factor, and $E_{i}=\left\langle \omega_{i}\right\rangle =\sum\omega_{i}p_{i}$
is the internal energy of $\omega_{i}$. Since the dark information
without dark energy is shown to be $I_{0}\left(a,b\right)=8\pi E_{a}E_{b}$
\cite{key-QYC}, we can conclude that the dark energy will increase
the dark information of the Hawking radiation. It can be seen from
Eq. (\ref{eq:I}) that when the cosmological constant $\Lambda\rightarrow0$,
the modification by the dark energy on the dark information vanishes,
i.e. , $\mu\rightarrow0$, then the result is naturally back to the
dark information $I_{0}$ in the case without dark energy. We can
further use the difference between $I$ and $I_{0}$ to measure the
amount of the dark information added by dark energy, which reads

\begin{equation}
\Delta I_{\Lambda}\left(a,b\right)\equiv I-I_{0}=32\pi\left(\xi+1\right)\left(2\xi+1\right)M^{2\xi}\left(\frac{2\Lambda}{3}\right)^{\xi}E_{a}E_{b}.
\end{equation}
In the cosmological constant model of dark energy,$\xi=1$, we have

\begin{equation}
I\left(a,b\right)=8\pi\left(1+16M^{2}\Lambda\right)E_{a}E_{b},
\end{equation}
so that

\begin{equation}
\Delta I_{\Lambda}\left(a,b\right)=128\pi M^{2}\Lambda E_{a}E_{b}.\label{eq:deltaI}
\end{equation}
Equation (\ref{eq:deltaI}) tells us that the dark information of
the black hole radiation added by the dark energy increases as the
mass of the black hole increases. When we pay attention to the Hawking
radiation process, we can infer that with the evaporation of black
hole, the black hole mass is decreasing, i.e. , $M\downarrow$. As
a result, the dark information added by dark energy is also decreasing,
i.e. , $\Delta I_{\Lambda}\left(a,b\right)\downarrow$, for the case
with $E_{a}$ and $E_{b}$ remain unchanged.

The results we obtained in this section show that the dark energy
not only has a gravitational effect, but also has an influence on
the information. 

\section{Remarks and conclusion }

In this paper, we revealed various influences of the dark energy on
the black hole and its radiation. Firstly, a general canonical typicality
based approach is developed to derive the non-thermal black hole radiation
spectrum without referring the dynamics of the particle tunneling.
This statistical mechanical approach only need to know the horizon
radius as the specific expression of the three hairs of the black
hole (mass, charge, angular momentum). With the Schwarzschild black
hole as an illustration, we calculated the black hole radiation spectrum
with the existence of the dark energy. The result is confirmed with
the ``standard'' approach\textemdash the quantum tunneling based
approach. From the corrected radiation spectrum, the dark energy is
showed to make the black hole radiation colder as given by Eq. (\ref{eq:HT}),
and the lifetime of black hole longer in Eq. (\ref{eq:t}). Moreover,
according to the black hole thermodynamics, we put forward the dark
energy based cooling mechanism (DECM) to explain the decreasing of
the Hawking temperature physically. When the non-thermal effect of
the black hole radiation is taken into account, we found that the
dark energy can increase the dark information of the black hole radiation,
as shown in Eq. (\ref{eq:deltaI}). This dark information is introduced
to describe the locally unobservable correlation among the black hole
radiations. 

Our investigation in this paper can be extended to other types of
black holes to find their radiation spectra as well as the influences
of the dark energy on their radiation. To this end, the follows two
steps are needed: (i) obtain the metric of the black hole with the
existence of dark energy, and then use it to derive the corrected
horizon radius; (ii) use the canonical typicality based approach to
get the black hole radiation spectrum. Once the radiation spectrum
is obtained, all the thermodynamic properties of the black hole radiation
field can be obtained from it. This study shows that the dark energy
has an influence on information (for black hole radiation, the dark
information increases), not just the mechanical effect of accelerating
the expansion of the universe. 

\paragraph{Acknowledgements}

We thank Qing-Yu Cai, Hui Dong, Xin Wang, and Shan-He Su for helpful
discussions. We also thank Hong-Jian He for drawing our attention
on his recent research about probing dark energy (Ref \cite{dark energy metric}).
This study is supported by the National Basic Research Program of
China (Grant No. 2016YFA0301201 \& No. 2014CB921403), the NSFC (Grant
No. 11534002), and the NSAF (Grant No. U1730449 \& No. U1530401) 

\paragraph{Note added}

After we finished this study and the paper was completed, we note
that the Hawing temperature in the case with dark energy has also
been discussed in Ref \cite{black hole as heat engine}, where the
authors treated a charged black hole as a heat engine and investigated
the effect that the dark energy on the heat engine's performance.
The result they obtained seems consistent with the dark energy based
cooling mechanism we presented.

The universe U, which contains the system of interest B (e.g. , black
hole) and the environment O, is assumed to be in an arbitrary pure
state

\begin{equation}
\left|\Psi\right\rangle =\sum_{b}\sum_{o}\frac{C\left(b,o\right)}{\sqrt{\Omega_{\textrm{U}}}}\left|b\right\rangle \otimes\left|o\right\rangle ,\label{eq:state}
\end{equation}
where $\Omega_{\textrm{U}}=\Omega_{\textrm{U}}\left(E_{\textrm{U}}\right)$
is the total number of microstates for the universe with energy $E_{\textrm{U}}$,
and $C\left(b,o\right)$ is the coefficient of state $\left|b\right\rangle \otimes\left|o\right\rangle $.
And $\left|b\right\rangle $ and $\left|o\right\rangle $ are the
eigen-states of B and O, respectively. The orthogonal conditions are
$\left\langle b_{i}|b_{j}\right\rangle =\delta_{ij}$, and $\left\langle o_{k}|o_{l}\right\rangle =\delta_{kl}$.
Using the normalization condition of state $\left|\Psi\right\rangle $,
we have

\begin{equation}
\frac{\sum_{b}\sum_{o}\left|C\left(b,o\right)\right|^{2}}{\Omega_{\textrm{U}}}=1,
\end{equation}
which means the average value of $\left|C\left(b,o\right)\right|^{2}$

\begin{equation}
\left\langle \left|C\left(b,o\right)\right|^{2}\right\rangle =1.
\end{equation}
Let $\left|\Psi_{b}\right\rangle =\sum_{o}C\left(b,o\right)\left|o\right\rangle $.
Then the universe state can be re-expressed as

\begin{equation}
\left|\Psi\right\rangle =\frac{\sum_{b}\left|b\right\rangle \left|\Psi_{b}\right\rangle }{\sqrt{\Omega_{\textrm{U}}}}.
\end{equation}
As a result, by tracing over the environment O, the reduced density
matrix of B is obtained as

\begin{equation}
\rho_{\textrm{B}}=\textrm{Tr}_{\textrm{O}}\left(\left|\Psi\right\rangle \left\langle \Psi\right|\right)=\frac{1}{\Omega_{\textrm{U}}}\sum_{b_{i},b_{j}}\left\langle \Psi_{b_{i}}|\Psi_{b_{j}}\right\rangle \left|b_{i}\right\rangle \left\langle b_{j}\right|,\label{eq:rouBa}
\end{equation}
where $\left\langle \Psi_{b_{i}}|\Psi_{b_{j}}\right\rangle =\delta_{ij}\left\langle \Psi_{b}|\Psi_{b}\right\rangle $
\cite{key-CT2}. For the environment O supported in a high-dimension
Hilbert space, according to central limit theorem, we have

\begin{equation}
\left\langle \Psi_{b}|\Psi_{b}\right\rangle =\sum_{o}\left|C\left(b,o\right)\right|^{2}=\Omega_{\textrm{O}}\left(E_{\textrm{U}}-E_{b}\right),
\end{equation}
where $\Omega_{\textrm{O}}\left(E\right)$ is the number of O's micro-states
with energy $E=E_{\textrm{U}}-E_{b}$. The reduced density matrix
of B is thus simplified as

\begin{equation}
\rho_{\textrm{B}}=\sum_{b}\frac{\Omega_{\textrm{O}}\left(E_{\textrm{U}}-E_{b}\right)}{\Omega_{\textrm{U}}}\left|b\right\rangle \left\langle b\right|.
\end{equation}
For B is a macroscopic object, the eigen-energy of its macrostate
$\left|b\right\rangle $ can be approximated as $E_{b}\approx E$,
with the fluctuation being about $\Delta E_{b}\sim E/N$, where $N$
is the micro - degrees of freedom of B. Obviously, the energy fluctuation
vanishes in the thermodynamic limit, i.e, $N\rightarrow\infty$, thus
we have

\begin{equation}
\rho_{\textrm{B}}=\frac{\sum_{b}\left|b\right\rangle \left\langle b\right|}{\Omega_{\textrm{B}}\left(E\right)},
\end{equation}
with

\begin{equation}
\Omega_{\textrm{B}}\left(E\right)\equiv\frac{\Omega_{\textrm{U}}}{\Omega_{\textrm{O}}\left(E_{\textrm{U}}-E\right)}.
\end{equation}
This means that B obeys the micro-canonical distribution. When look
at B's subsystem R, the rest of B is denoted as $\mathrm{B'=B-R}$.
The reduced density matrix of B can be rewritten as

\begin{equation}
\rho_{\textrm{B}}=\frac{\sum_{r,b'}\left|r,b'\right\rangle \left\langle r,b'\right|}{\Omega_{\textrm{B}}\left(E\right)},
\end{equation}
where$\left|r\right\rangle $ and $\left|b'\right\rangle $ are the
eigenstates of R and $\textrm{B}'$ with eigen-energies $E_{r}$ and
$E_{b'}$, respectively. $E_{r}+E_{b'}=E$ is the constraint condition
given by energy conservation . The reduced density matrix of R

\begin{equation}
\rho_{\textrm{R}}=\textrm{Tr}{}_{\textrm{B}'}\left(\rho_{\textrm{B}}\right)=\sum_{r}\frac{\Omega_{\textrm{B}'}\left(E-E_{r}\right)}{\Omega_{\textrm{B}}\left(E\right)}\left|r\right\rangle \left\langle r\right|,\label{eq:densityR}
\end{equation}
is obtained by tracing over $B'$. Here, $\Omega_{\textrm{B}'}\left(E-E_{r}\right)$
is the number of micro-states of the system $\textrm{B}'$ with energy
$E-E_{r}$. Making use of the Boltzmann entropy

\begin{equation}
S_{\textrm{B}}\left(E\right)=\ln\left[\Omega_{\textrm{B}}\left(E\right)\right],
\end{equation}
and

\begin{equation}
S_{\textrm{B}'}\left(E-E_{r}\right)=\ln\left[\Omega_{\textrm{B}'}\left(E-E_{r}\right)\right],
\end{equation}
Eq. (\ref{eq:densityR}) can be further written as

\begin{equation}
\rho_{\textrm{R}}=\textrm{Tr}{}_{\textrm{B}'}\left(\rho_{\textrm{B}}\right)=\sum_{r}\textrm{e}^{-\Delta S_{\textrm{BB}'}\left(E_{r},E\right)}\left|r\right\rangle \left\langle r\right|,\label{eq:densityRa-1}
\end{equation}
where $\Delta S_{\textrm{BB}'}\left(E_{r},E\right)\equiv S_{\textrm{B}}\left(E\right)-S_{\textrm{B}'}\left(E-E_{r}\right)$
is the difference in entropy between B and $\textrm{B}'$ . We can
clearly see from Eq. (\ref{eq:densityRa-1}) that only when $\Delta S_{\textrm{BB}'}$
is linearly dependent on $E_{r}$, the spectrum of R is perfectly
thermal. What should be mentioned here is that we do not expand $\Delta S_{\textrm{BB}'}$
only up to the first order of $E_{r}$, as done in the most studies
in the thermodynamic limit. 

Now we apply the above result to black holes. For an arbitrary black
hole, there are three macro parameters to describes its geometry and
number of microstate. These parameters, known as hairs, are mass $M$,
charge $Q$, and angular momentum $J$. Considering the conservation
laws for hairs, we can generalize Eq. (\ref{eq:densityR}) as

\begin{equation}
\rho_{\textrm{R}}=\sum_{\omega,q,j}\frac{\Omega_{\textrm{B}'}\left(M-\omega,Q-q,J-j\right)}{\Omega_{\textrm{B}}\left(M,Q,J\right)}\left|\omega,q,j\right\rangle \left\langle \omega,q,j\right|,
\end{equation}
where $\left|\omega,q,j\right\rangle $ is the eigenstate of R with
mass $\omega$, charge $q$, and angular momentum $j$. Then, by expressing
the number of microstate with entropy, and making use of the Bekenstein-Hawking
(B-H) entropy for B and B'

\begin{equation}
S_{\textrm{B}}\left(M,Q,J\right)=S_{BH}\left(M,Q,J\right)=\frac{A_{H}\left(M,Q,J\right)}{4}=\pi R_{H}^{2}\left(M,Q,J\right),
\end{equation}
and

\begin{equation}
S_{\textrm{B}}\left(M-\omega,Q-q,J-j\right)=S_{BH}\left(M-\omega,Q-q,J-j\right)=\pi R_{H}^{2}\left(M-\omega,Q-q,J-j\right),
\end{equation}
we have 

\begin{equation}
\rho_{\textrm{R}}=\sum_{\omega,q,j}\exp\left[\pi R_{H}^{2}\left(M-\omega,Q-q,J-j\right)-\pi R_{H}^{2}\left(M,Q,J\right)\right]\left|\omega,q,j\right\rangle \left\langle \omega,q,j\right|.
\end{equation}
Here, $A_{H}\left(M,Q,J\right)$ and $R_{H}\left(M,Q,J\right)$ is
the area and radius of the horizon, respectively. Therefore, we obtain
the distribution probability of $\left|\omega,q,j\right\rangle $

\begin{equation}
p\left(\omega,q,j,M,Q,J\right)=\textrm{e}^{-\pi\left[R_{H}^{2}\left(M,Q,J\right)-R_{H}^{2}\left(M-\omega,Q-q,J-j\right)\right]}.
\end{equation}

\section{Dark information of black hole radiation}

From Eq. (\ref{eq:p}), one has the distribution possibility for particles
$a$ and $b$ as

\begin{equation}
p_{a}=p\left(\omega_{a},M\right)=\textrm{exp}\left(-\beta_{H}\omega_{a}+\chi\omega_{a}^{2}\right),\label{eq:pa}
\end{equation}

\begin{equation}
p_{b}=p\left(\omega_{b},M\right)=\textrm{exp}(-\beta_{H}\omega_{b}+\chi\omega_{b}^{2}),\label{eq:pb}
\end{equation}
and the joint possibility

\begin{equation}
p_{a,b}=p\left(\omega_{a}+\omega_{b},M\right)=\textrm{exp}\left[-\beta_{H}\left(\omega_{a}+\omega_{b}\right)+\chi\left(\omega_{a}+\omega_{b}\right)^{2}\right],\label{eq:pab}
\end{equation}
where the inverse radiation temperature

\begin{equation}
\beta_{H}=8\pi M\left[1+4\left(\xi+1\right)M^{2\xi}f\right],
\end{equation}
and the second order coefficient 

\begin{equation}
\chi=4\pi\left[1+4\left(\xi+1\right)\left(2\xi+1\right)M^{2\xi}f\right].
\end{equation}
Following from Eqs. (\ref{eq:pa}), (\ref{eq:pb}), and (\ref{eq:pab}),
we find

\begin{equation}
\frac{p_{a,b}}{p_{a}p_{b}}=\textrm{e}^{2\chi\omega_{a}\omega_{b}}\neq0,
\end{equation}
which implies that these two radiated particles have correlation.
Substituting the above equation into Eq. (\ref{eq:I-1}) and then
the dark information is obtained as

\begin{equation}
I\left(a,b\right)=2\chi\sum_{a,b}p_{a,b}\omega_{a}\omega_{b}.
\end{equation}
Note that the joint possibility satisfies $p_{a,b}=p\left(\omega_{a},M\right)p\left(\omega_{b},M-\omega_{a}\right)$,
thus, with the replacement $M-\omega_{a}\rightarrow M'$, 

\begin{equation}
I\left(a,b\right)=2\chi\left[\sum_{\omega_{a}=0}^{\omega_{a}=M}p\left(\omega_{a},M\right)\omega_{a}\right]\left[\sum_{\omega_{b}=0}^{\omega_{b}=M'}p\left(\omega_{b},M'\right)\omega_{b}\right]
\end{equation}
Here,

\begin{equation}
\left[\sum_{\omega_{a}=0}^{\omega_{a}=M}p\left(\omega_{a},M\right)\omega_{a}\right]=\left\langle \omega_{a}\right\rangle \equiv E_{a}
\end{equation}
and 

\begin{equation}
\left[\sum_{\omega_{b}=0}^{\omega_{b}=M'}p\left(\omega_{b},M'\right)\omega_{b}\right]=\left\langle \omega_{b}\right\rangle \equiv E_{b}
\end{equation}
are the internal energy of particles $a$ and $b$, respectively.
Finally, we obtain the dark information of black hole radiation with
the correction of dark energy as

\begin{equation}
I\left(a,b\right)=2\chi E_{a}E_{b}=8\pi\left[1+4\left(\xi+1\right)\left(2\xi+1\right)M^{2\xi}f\right]E_{a}E_{b},
\end{equation}
this is what we illustrated in Eq. (\ref{eq:I}).
\end{document}